\documentclass[twocolumn]{aastex62}

\usepackage{amsmath}
\usepackage{color}
\graphicspath{{./}{figures/}}
\usepackage{natbib}
\usepackage{enumitem}
\usepackage{float}
\usepackage{url}
\usepackage{color,soul}

\journalinfo{The Astrophysical Journal Letters}
\submitjournal{ApJ Letters}

\shorttitle{TRGB Distance to NGC\,1052-DF4}
\shortauthors{Danieli et al.}

\begin{document}

\gdef\kms{km\,s$^{-1}$}

\title{A Tip of the Red Giant Branch Distance to the Dark Matter Deficient Galaxy NGC\,1052-DF4 from Deep {\em Hubble Space Telescope} Data}

\correspondingauthor{Shany Danieli}
\email{shany.danieli@yale.edu, shanyi1@gmail.com}

\author[0000-0002-1841-2252]{Shany Danieli}
\affil{Department of Physics, Yale University, New Haven, CT 06520, USA \\}
\affil{Yale Center for Astronomy and Astrophysics, Yale University, New Haven, CT 06511, USA \\}
\affil{Department of Astronomy, Yale University, New Haven, CT 06511, USA \\}

\author[0000-0002-8282-9888]{Pieter van Dokkum}
\affiliation{Department of Astronomy, Yale University, New Haven, CT 06511, USA \\}

\author[0000-0002-4542-921X]{Roberto Abraham}
\affiliation{Department of Astronomy and Astrophysics, University of Toronto, Toronto ON, M5S 3H4, Canada\\}
\affiliation{Dunlap Institute for Astronomy and Astrophysics, University of Toronto, Toronto ON, M5S 3H4, Canada\\}

\author[0000-0002-1590-8551]{Charlie Conroy}
\affiliation{Harvard-Smithsonian Center for Astrophysics, 60 Garden Street, Cambridge, MA, USA\\}

\author[0000-0002-8084-8612]{Andrew E. Dolphin}
\affiliation{Raytheon, 1151 E. Hermans Road, Tuscon, AZ 85706, USA\\}
\affiliation{Steward Observatory, University of Arizona, 933 North Cherry Avenue, Tucson, AZ 85721-0065 USA}

\author[0000-0003-2473-0369]{Aaron J. Romanowsky}
\affiliation{Department of Physics and Astronomy, San Jos\'e State University, San Jose, CA 95192, USA\\}
\affiliation{University of California Observatories, 1156 High Street,
Santa Cruz, CA 95064, USA\\}


\begin{abstract}

Previous studies have shown that the large, diffuse galaxies NGC\,1052-DF2 and NGC\,1052-DF4 both have populations of unusually luminous globular clusters as well as a very low dark matter content. Here we present newly-obtained deep {\em Hubble Space Telescope} ({\em HST}) Advanced Camera for Surveys (ACS) imaging of one of these galaxies, NGC\,1052-DF4. We use these data to measure the distance of the galaxy from the location of the tip of the red giant branch (TRGB). We find a rapid increase in the number of detected stars fainter than $m_{F814W} \sim 27.3$, which we identify as the onset of the red giant branch. Using a forward modeling approach that takes the photometric uncertainties into account, we find a TRGB magnitude of $m_{F814W,\rm TRGB}=27.47 \pm 0.16$. The inferred distance, including the uncertainty in the absolute calibration, is $D_{\rm TRGB}=20.0 \pm 1.6$\,Mpc. The TRGB distance of NGC\,1052-DF4 is consistent with the previously-determined surface brightness fluctuation distance of $D_{\rm SBF}=18.7\pm 1.7$\,Mpc to NGC\,1052-DF2 and is consistent with the distance of the bright elliptical galaxy NGC\,1052. We conclude that the unusual properties of these galaxies cannot be explained by distance errors.

\end{abstract}


\section{Introduction} \label{sec:intro}

In 2016, as part of the Dragonfly Nearby Galaxies Survey (DNGS; \citealt{2016ApJ...830...62M}), we identified 23 low surface brightness galaxies in four survey fields: NGC\,1052, NGC\,1084, M\,96, and NGC\,4258. Follow-up {\em HST}/ACS observations (\citealt{2018ApJ...868...96C}) revealed that two of these galaxies, NGC\,1052-DF2 and NGC\,1052-DF4, in the field of the elliptical galaxy NGC\,1052, have unusual properties with respect to the rest of the sample. At the distance of NGC\,1052 ($D_{1052}=19.4-21.4$\,Mpc; \citealt{2001ApJ...546..681T}; \citealt{2002MNRAS.330..443B}), NGC\,1052-DF2 and NGC\,1052-DF4 have the stellar masses of dwarf galaxies ($\sim 2 \times 10^8 \ \mathrm{M}_{\odot}$ and $\sim 1.5 \times 10^8 \ \mathrm{M}_{\odot}$, respectively) but large sizes ($R_e=2.2$\,kpc and $1.6$\,kpc). Furthermore, they both host a spectacular population of luminous globular clusters. The specific frequency and median luminosity of the clusters are much higher than seen in other galaxies with the same stellar mass (\citealt{2018ApJ...856L..30V}, \citeyear{2019ApJ...874L...5V}).

Unusual as these aspects are, their kinematics turned out to be even more surprising.
In 2018 we found that NGC\,1052-DF2 seems to be lacking in dark matter (\citealt{2018Natur.555..629V}), based on the radial
velocities of ten of its globular clusters. The inferred velocity dispersion of the galaxy is $\sigma_{\mathrm{gc}}=7.8^{+5.2}_{-2.2} \ \mathrm{kms}^{-1}$ (\citealt{2018RNAAS...2b..54V}). 
Some studies raised concerns about these results, largely due to the small number of tracers (\citealt{2018ApJ...859L...5M}; \citealt{2019MNRAS.484..245L}). Follow-up spectroscopy of the diffuse stellar light with the Keck Cosmic Web Imager (KCWI) and the Multi-Unit Spectroscopic Explorer (MUSE) confirmed the low dispersion of NGC\,1052-DF2 (\citealt{2019ApJ...874L..12D}; \citealt{2019A&A...625A..76E}). In particular, the high resolution KCWI data gave $\sigma_{\mathrm{stars}}=8.5^{+2.3}_{-3.1} \ \mathrm{kms}^{-1}$, fully consistent with the results inferred from the globular clusters (\citealt{2019ApJ...874L..12D}). 
Similarly, NGC\,1052-DF4 has an inferred velocity dispersion of $\sigma_{\mathrm{gc}}=4.2^{+4.4}_{-2.2} \ \mathrm{kms}^{-1}$ based on seven globular clusters (\citealt{2019ApJ...874L...5V}), consistent with the expected value from the stars alone ($7 \ \mathrm{kms}^{-1}$). 
Given the spatial extent of these galaxies, we do not expect them to be baryon-dominated within the
effective radius (see, e.g., \citealt{2019ApJ...880...91V}).
The two galaxies share essentially all their anomalous properties: they have similar sizes, luminosities, and colors; the same morphology; they both have a population of luminous globular clusters; and they both have a velocity dispersion that is consistent with that expected from the stellar mass alone.

Many of the unusual properties of NGC\,1052-DF2 and NGC\,1052-DF4 are distance-dependent, and accurate constraints on the dark matter content and the globular cluster luminosity function require accurate distances. The canonical distance to both galaxies is $\sim 19 \ \mathrm{Mpc}$, based on their radial velocities, their presumed membership of the NGC\,1052 group, and a surface brightness fluctuation (SBF) measurement from the then-available 2-orbit (1 orbit
$F814W$ and 1 orbit $V_{606}$) HST data (\citealt{2018ApJ...864L..18V}; \citealt{2018RNAAS...2c.146B}; \citealt{2018ApJ...868...96C}). In particular, van Dokkum et al.\ (\citeyear{2018ApJ...864L..18V}) use a megamaser-TRGB-SBF distance ladder that is free of calibration uncertainties to determine a distance of $D_{\rm SBF}=18.7 \pm 1.7 \ \mathrm{Mpc}$ to NGC\,1052-DF2. However, others have suggested that the galaxies are significantly closer to us.
Trujillo et al.\ \citeyear{2019MNRAS.486.1192T} and Monelli et al.\ \citeyear{2019ApJ...880L..11M} derive distances of 
$\approx 13 \ \mathrm{Mpc}$ and $14.2\pm 0.7 \ \mathrm{Mpc}$ to NGC\,1052-DF2 and NGC\,1052-DF4 respectively, from the same HST imaging. In contrast to van Dokkum et al.\ (\citeyear{2018ApJ...864L..18V}) they associate faint detections in the HST data with individual stars on the red giant branch, and associate the brightest of these detections with its tip (TRGB).  A shorter distance would bring NGC\,1052-DF2 and NGC\,1052-DF4 more in line with other galaxies: their physical sizes would be smaller, their globular clusters less luminous, and their mass-to-light ratios slightly higher.\footnote{A short distance is not without problems: the radial velocities of the galaxies, their surface brightness fluctuation signal, and their apparent isolation in the foreground of the NGC\,1052 group would all require explanations.}

These disagreements can be resolved using deeper data.  The previously available 1+1 orbit HST data are not deep enough to detect the TRGB beyond $D\sim 15$\,Mpc, and the color-magnitude distributions (CMDs) generated by the two studies are highly sensitive to the quality cuts that are applied to the photometry. 
In this \textit{Letter} we present significantly deeper {\em HST}/ACS observations of NGC\,1052-DF4, obtained in a Cycle 26 mid-cycle program. Using a total of 8 orbits in $F814W$ and 4 orbits in $F606W$ we measure the distance to the galaxy based on a secure determination of the TRGB magnitude. Vega magnitudes are used throughout this paper.

\begin{figure*}[p]
{\centering
  \includegraphics[width=1.0\textwidth]{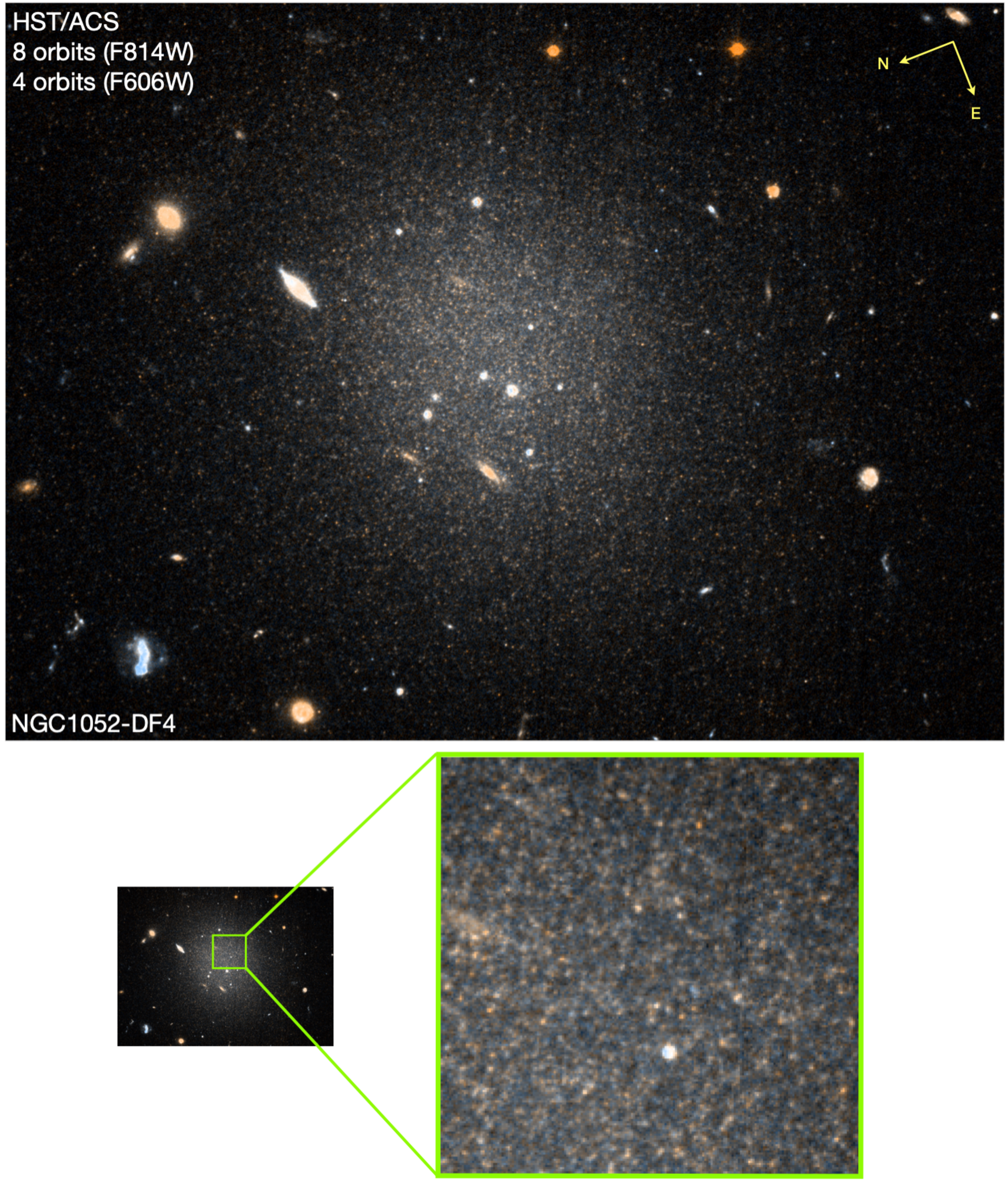}
  \caption{Deep {\em HST} image of NGC\,1052-DF4 obtained with the ACS. The total exposure time is $16,760$ seconds (8 orbits) in the  $F814W$ filter and $8,240$ seconds (4 orbits) in the $F606W$ filter. The bottom panel shows a zoom on the central region of the galaxy. The image reveals a resolved red giant branch against the glow of unresolved bluer subgiants and main sequence stars.}
  \label{fig:colorimage}
}
\end{figure*}


\section{{\em HST}/ACS Observations and Processing} \label{sec:observations}

NGC\,1052-DF4 was observed with the {\em HST} ACS Wide-Field Channel (WFC) in Cycle 26 mid-cycle program 15695 in July 2019 in three visits. We obtained seven orbits in the $F814W$ filter and three orbits in the $F606W$ filter. Adding these data to the observations that we obtained in 2017 (program 14644), the total exposure time is $16,760$\,s in $F814W$ and $8,240$\,s in $F606W$. The pointing and orientation were constrained to ensure that the neighboring galaxy NGC\,1052-DF5 (with a projected distance of $\sim 90\arcsec$ from NGC\,1052-DF4) would be included in the field of view. NGC\,1052-DF4 was placed near the center of one of the WFC chips.

The observations were obtained with a standard dither pattern with four exposures in each orbit to optimally sample the point spread function and to facilitate the identification of hot pixels and cosmic rays. Individual exposures were bias-subtracted and dark-current-subtracted, flat-fielded, and CTE-corrected by the STScI pipeline, providing calibrated \texttt{flc} files. We used the \texttt{TweakReg} utility to align the 48 \texttt{flc} files and AstroDrizzle to combine the aligned \texttt{flc} images into astrometrically-corrected drizzled images (\texttt{drc}) in the $F814W$ and $F606W$ filters.

A color image of NGC\,1052-DF4, generated from the drizzled images, is shown in Figure \ref{fig:colorimage}. As can be seen in the zoomed-in panel, the galaxy is well-resolved into stars in these deep data. The red giants appear as thousands of yellow point sources against a bluer unresolved backdrop of subgiants and main sequence stars. The much-fainter neighboring galaxy NGC\,1052-DF5 is not shown, but we note here that it is also resolved into stars: approximately 200 giants are detected down to $m_{F814W}<28$ mag within two effective radii \citep[with $R_e$ from][]{2018ApJ...868...96C}. This enigmatic object is the topic of a future paper.

\section{Stellar Photometry} \label{sec:photometry}

\begin{figure*}[t!]
{\centering
  \includegraphics[width=0.9\textwidth]{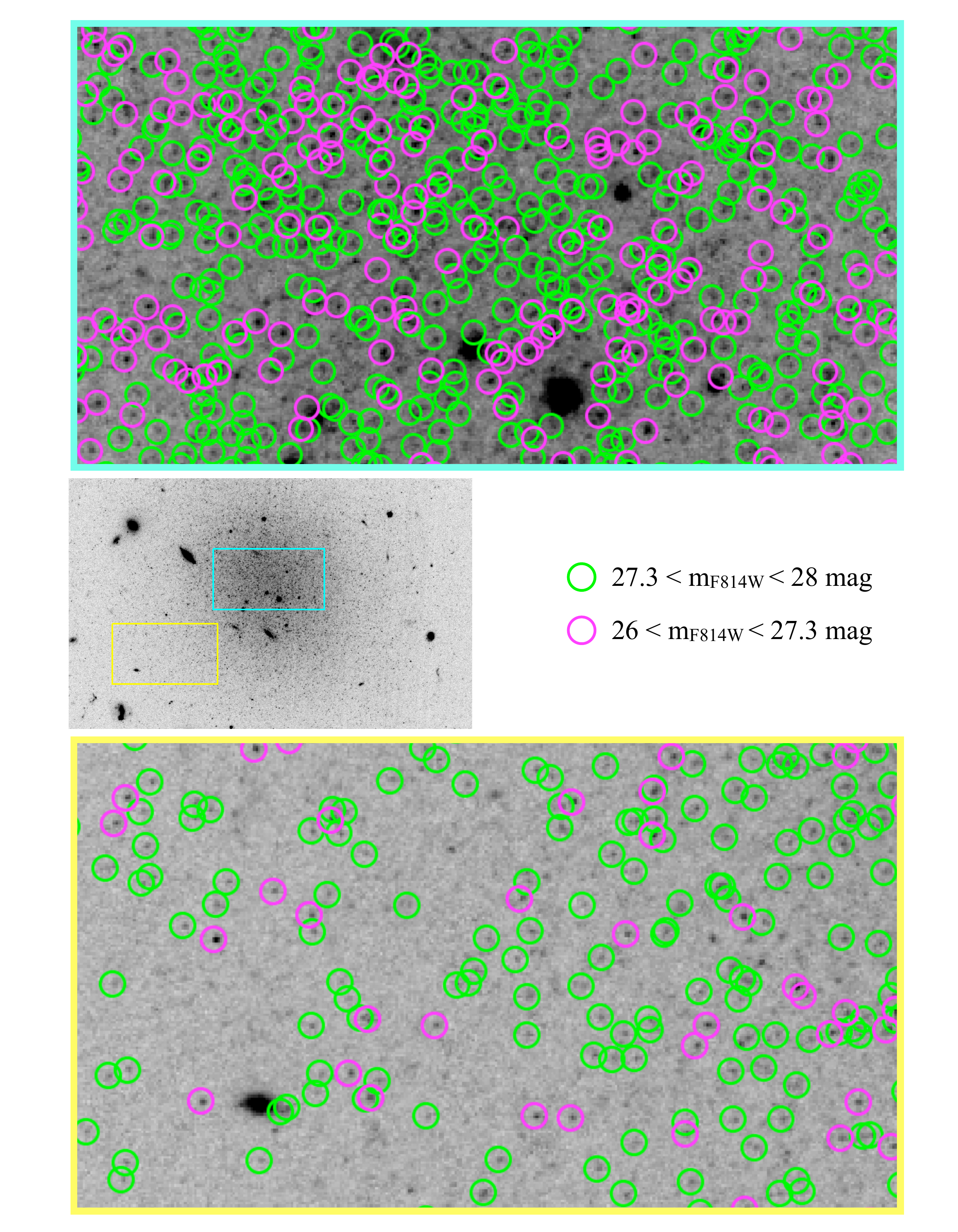}
  \caption{Detected stars (after quality cuts) from the stellar photometry catalog, shown in small regions in the outskirts (yellow) and in the center (cyan) of NGC\,1052-DF4. RGB stars with $27.3<m_{F814W}<28 \ \mathrm{mag}$ are circled in green and likely AGB stars with $26<m_{F814W}<27.3 \ \mathrm{mag}$ in magenta. Note that many objects do not survive the quality cuts, particularly in the center where blends dominate over isolated stars. Stars in the central part of the galaxy ($R<R_{\mathrm{eff}}$) are excluded from the TRGB measurement.}
  \label{fig:photometry}
}
\end{figure*}

Photometry was carried out on the bias subtracted, flat-fielded, CTE-corrected \texttt{flc} images, produced by the STScI ACS pipeline. We used the software package DOLPHOT\footnote{\url{http://americano.dolphinsim.com/dolphot/}}, which is a modified version of HSTphot (\citealt{2000PASP..112.1383D}). The DOLPHOT/ACS package includes a sequence of image preparation steps that are performed prior to the detection of stars. These include masking of bad columns and hot pixels (\texttt{acsmask}) and background determination using the \texttt{calcsky} module. These pre-processing steps were run on the 48 \texttt{flc} files as well as on the chosen reference \texttt{drc} image (the deepest drizzled image was chosen with 32 stacked \texttt{flc} images in $F814W$ and a total exposure time of 16,760\,s). 

DOLPHOT detects stars by fitting PSFs generated with Tiny Tim (\citealt{1995ASPC...77..349K}) simultaneously to all 48 \texttt{flc} images. This approach has the advantage that no resampling of the data is required and the noise properties of the observations are conserved. After detection in the full stack the software measures fluxes separately in the $F814W$ and $F606W$ frames.
The three DOLPHOT parameters that have the strongest influence on the photometry are the sky fitting parameter (\texttt{FitSky}), the aperture radius (\texttt{RAper}) and the \texttt{Force1} parameter which forces all sources detected to be fitted as stars. We adopted the following: \texttt{FitSky}=2, \texttt{RAper}=3 and \texttt{Force1}=1. Those are similar to the recommended parameters on the DOLPHOT website,  those used in the ANGST survey (\citealt{2009ApJS..183...67D}), and in the GHOSTS project (\citealt{2011ApJS..195...18R}) and optimize the photometry for cases of crowded stellar fields. As discussed in \S\,\ref{sec:forward}, with these parameters we find a systematic error of $0.1 \ \mathrm{mag}$ at $m_{F814W}=27.0$ in the recovered magnitudes of artificial stars.

The raw DOLPHOT output was corrected for Galactic extinction of 0.041\,mag in $F814W$ and $0.066$\,mag in $F606W$ (\citealt{2011ApJ...737..103S}), and filtered to isolate stars with reliable photometry. Only objects with signal-to-noise ratio $>4$ in $F814W$, signal-to-noise ratio $>2$ in $F606W$, object-type $\leq 2$ (that is, {\em good star} or {\em faint star}), $|\mathrm{sharp}_{F814W}| \leq 0.5$,  $|\mathrm{sharp}_{F606W}| \leq 0.5$,  $\mathrm{crowd}_{F814W} \leq 0.5$,and $\mathrm{crowd}_{F606W} \leq 0.5$ were retained.
We also obtained photometry of 200,000 artificial stars uniformly covering the magnitude and color ranges $25<F814W<29$ and $1<F606W-F814W<1.5$. Further, the stars were distributed between $R_{\mathrm{eff}}$ and $4R_{\mathrm{eff}}$, i.e., excluding the central part of the galaxy where crowding is most severe. Stars were injected at random positions out to 4 effective radii into the \texttt{flc} frames, and treated in exactly the same way as the actual photometry.

In Figure \ref{fig:photometry} we show examples of stellar detections of small regions in the center and in the outskirts of the galaxy. With the new deep HST images, the signal-to-noise is just high enough to obtain photometry of stars down to $m_{F814W}\approx 28$ mag (see \S\,\ref{sec:trgb} for the uncertainty as a function of magnitude). In the central regions many objects are rejected as they are blended. In the outskirts we find a large population of unblended stars.

\section{Distance from the TRGB}\label{sec:trgb}
\subsection{Color-magnitude diagrams}

\begin{figure*}[t!]
{\centering
  \includegraphics[width=0.85\textwidth]{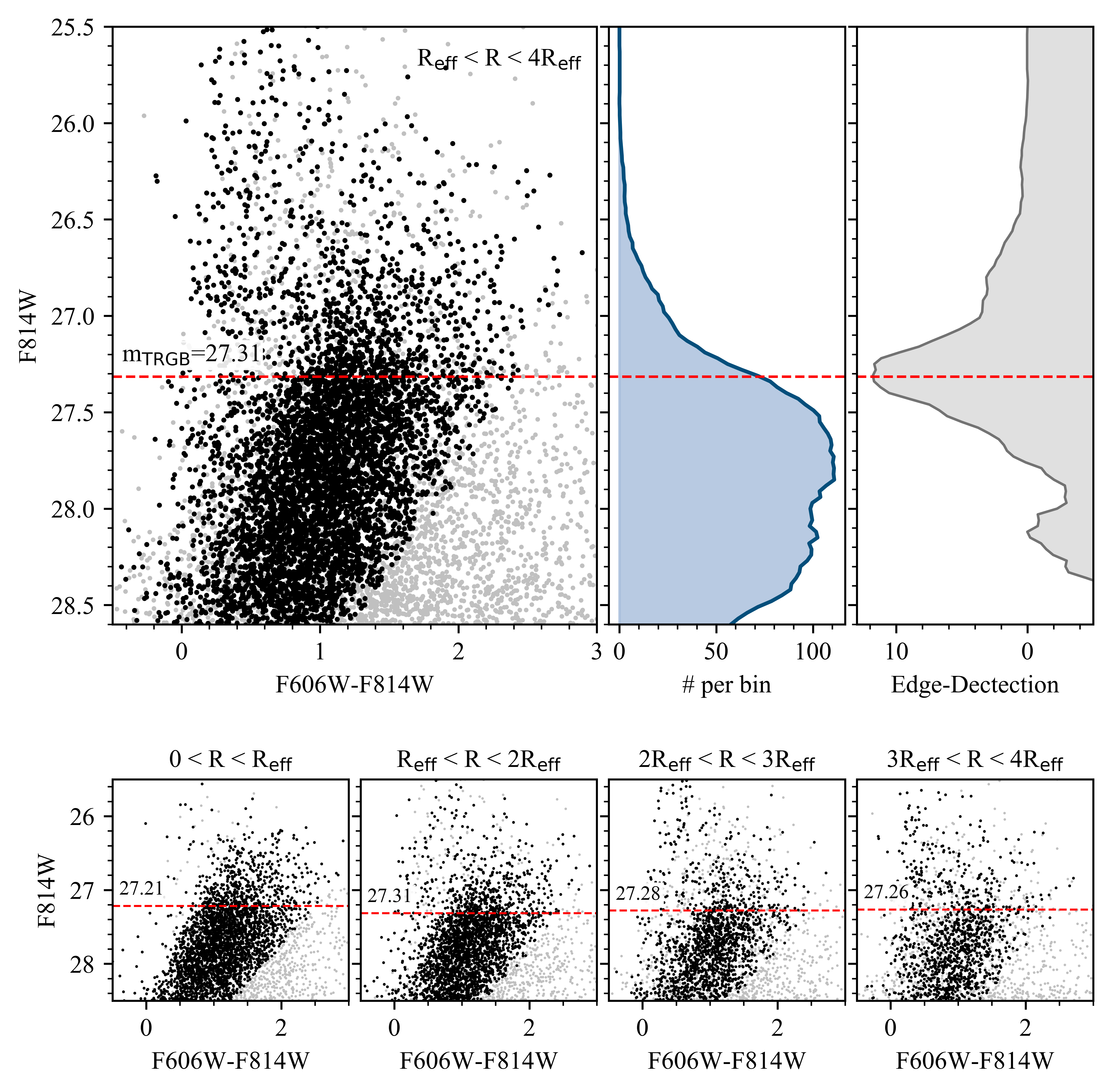}    
  \caption{TRGB edge detection for NGC\,1052-DF4. Top panels: the color-magnitude diagram (left), the binned luminosity function in $F814W$ in 0.03 mag intervals (middle) and the response function of the [-1,0,+1] kernel on the smoothed luminosity function, for stars within the radial bin $R_{\mathrm{eff}}<R<4R_{\mathrm{eff}}$.  The location of the measured TRGB magnitude is marked with the dashed red line where we measure a $F814W$ TRGB magnitude of $m_{F814W,\mathrm{TRGB}}=27.31 \pm 0.10 \ \mathrm{mag}$}. Bottom panels: the color-magnitude diagrams for various radial bins with the corresponding measured TRGB magnitude shown with the dashed red line. Black points show detected point sources remaining after applying quality cuts and light gray points show all detections before applying quality and color cuts to the photometry.
  \label{fig:cmdtrgb}
}
\end{figure*}

Color-magnitude diagrams (CMDs) for different radial bins are shown in the main panel and the bottom panels of Figure \ref{fig:cmdtrgb}. 
The CMDs show the characteristic color-magnitude trend of the bright end of the RGB (see, e.g., \citealt{2006AJ....132.2729M}; \citealt{2018ApJ...866..145H}). The scatter is driven by the large uncertainties in the $F606W$ magnitudes: the stars are red, and the exposure time in $F606W$ is only half that in $F814W$. These color uncertainties do not affect the $F814W$ luminosity function, which is used to derive the distance.
There is a characteristic increase in the number of stars fainter than $m_{F814W} \sim 27.3$ mag, which we identify as the onset of the red giant branch, i.e., the TRGB. This increase is illustrated by the difference between the number of magenta and green circles in Figure \ref{fig:photometry}.

The TRGB marks the core helium flash of first-ascent red giant branch stars.
Observationally, this physical phenomenon causes a sharp cutoff of the bright end of the red giant branch luminosity function, located at $M_{I, \mathrm{TRGB} } \approx -4.0$. In the following sections we describe the determination of the TRGB magnitude and the associated distance using two methods: a standard edge detection and forward modeling of the $F814W$-band stellar luminosity function of the galaxy.

\subsection{Edge detection}

The edge-detection method measures the first-derivative of a finely binned and smoothed luminosity function of the RGB and AGB populations. Briefly, a modified Sobel kernel, [$-1$,$0$,$+1$], is used as a filter and is applied to the Gaussian-smoothed (with a smoothing scale of 0.25 mags) $F814W$-band luminosity function. When photometric errors are small the method returns a maximum response at the location of the TRGB, where the slope of the luminosity function reaches a global maximum. The peak in the filtered luminosity function is fitted with a Gaussian to determine its location and width. The method has been used widely (e.g. \citealt{1993ApJ...417..553L}; \citealt{2006ApJ...651..822C}; \citealt{2015ApJ...800...13P}; \citealt{2015ApJ...812L..13S}; \citealt{2018A&A...615A..96M}; \citealt{2018ApJ...866..145H};  \citealt{2019arXiv190806120B}).

Figure \ref{fig:cmdtrgb} shows the results of the TRGB measurement using the edge-detection method. The main result was determined from a sample in the radial range $R_{\rm eff}<R<4R_{\rm eff}$ (upper panel), with the quality cuts of \S\,\ref{sec:photometry}. Blue stars with $(F814W-28)+1.5\times(F606W-F814W)>0.5$ are omitted. We measure a $F814W$ TRGB magnitude of $m_{F814W,\mathrm{TRGB}}=27.31 \pm 0.10 \ \mathrm{mag}$ for NGC\,1052-DF4, where the uncertainty is taken as the width of the Gaussian fit to the peak. The bottom panel of Figure \ref{fig:cmdtrgb} shows results from other radial bins. The results are consistent for all radial bins including the innermost one ($0<R<R_{\rm eff}$), where the data are severely crowded and most of the initial DOLPHOT detections are rejected by the quality cuts. This validates the quality cuts that are applied to the photometry and their ability to minimize crowding effects.

The uncertainty does not include systematic errors due to photometric errors or the calibration of the TRGB magnitude. As we show in \S\,\ref{sec:forward}, these uncertainties have a significant effect on the location of the TRGB.

\subsection{Forward modeling}\label{sec:forward}
\begin{figure*}[t!]
{\centering
  \includegraphics[width=1\textwidth]{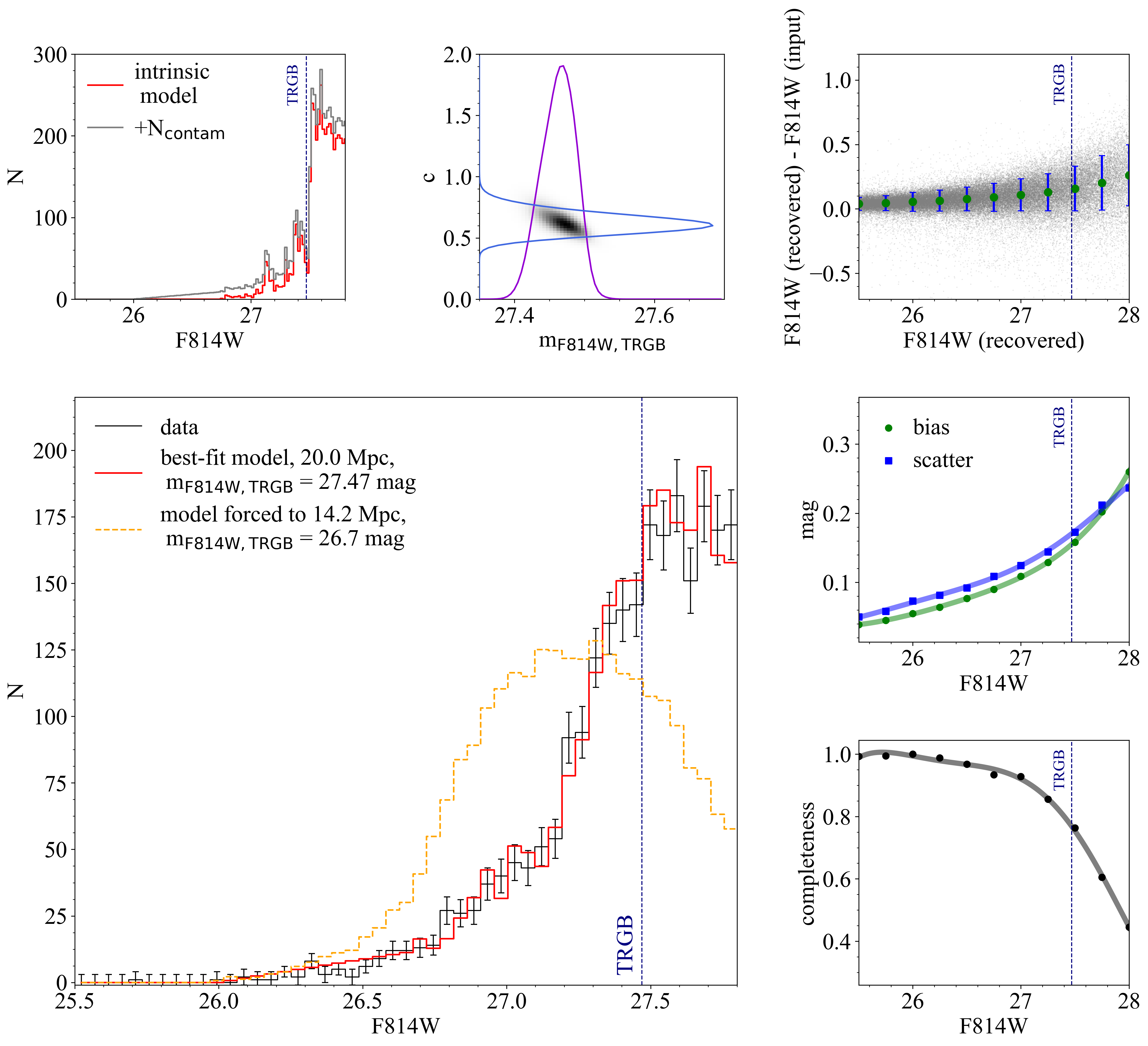}
  \caption{Results from the forward modeling fit to the $I$-band luminosity function. The top left panel shows a luminosity function with a TRGB magnitude of $27.47$\,mag. Panels on the right show the systematic offset (bias) and the scatter between true and recovered magnitudes using the artificial stars test (top and middle), as well as the completeness (bottom). The main panel shows the observed luminosity function (black), as well as the intrinsic model perturbed with the measurement errors (red). This model is an excellent fit. The likelihood is indicated in the central panel of the top row.}
  \label{fig:forward-modeling}
}
\end{figure*}

Photometric errors take three forms: systematic errors (``bias"), scatter, and incompleteness. Following previous studies (e.g. \citealt{2012ApJS..200...18D}; \citealt{2015ApJ...811..114G}; \citealt{2018ApJ...864..111C}; \citealt{2019arXiv190603230B}) we quantify all three aspects using the photometry of artificial stars (see \S\, 3). The results are shown in Figure \ref{fig:forward-modeling}. Bright stars are recovered without bias and with very small scatter, but the errors gradually rise toward fainter magnitudes. At $m_{F814W}=27.0$ the recovered magnitudes are systematically 0.10\,mag fainter than the input magnitudes, the $1\sigma$ scatter is $0.12$\,mag, and the completeness, determined from the ratio of the output magnitude distribution to the input magnitude distribution, is 92\,\%. These artificial stars tests results are quantified with fifth-order polynomial fits, shown by the solid lines.

With the errors thus parameterized we determine the TRGB distance through a forward-modeling procedure, as follows. First, a fully populated model for the galaxy is generated. Stars are drawn stochastically from a MIST isochrone (\citealt{2016ApJS..222....8D}; \citealt{2016ApJ...823..102C}) with an age of 10\,Gyr and a metallicity $[\mathrm{Fe/H}]=-1$, to match the color of the integrated stellar population as measured from the HST images. The results are not sensitive to the exact choice of stellar population parameters. The model contains $10^9$ stars down to $M_{F814W} = 11.3 \ \mathrm{mag}$. For an assumed TRGB magnitude a model observed $F814W$-band magnitude distribution is calculated, down to $m_{F814W}=30$ mag. Next, observational errors are applied to the model by shifting the brightness of each star by the parameterized bias, perturbing it by a random number drawn from a Gaussian distribution with a width that is given by the parameterized scatter. Possible mismatch between the number of AGB stars in the model and the true number, as well as contamination by blends, field stars, compact galaxies, and noise is taken into account with a linear function of the form $N_{\rm contam} = 20c \times (m-26)$ for $m>26$, and $N_{\rm contam} = 0$ for $m<26$, with $c$ a fit parameter. $N_{\rm contam}$ is added to the model and the observed brightness function is multiplied by the completeness. Each model is therefore characterized by two parameters, $m_{F814W,\rm TRGB}$ and $c$, and we determine the best fitting values for these parameters by minimizing $\chi^2$. The method is similar to that employed in the \texttt{TRGBTOOL} (\citealt{2006AJ....132.2729M}), but with fewer free parameters.\footnote{The \texttt{TRGBTOOL} has four free parameters: the distance, the slope of the counts on each side of the TRGB, and the strength of the TRGB.} We note that our fitting method is independent of the absolute calibration of the MIST models: we do not fit directly for a distance but we fit for the location of the discontinuity in the apparent magnitude distribution. This TRGB location is then converted into a distance modulus using the standard calibration in this field (see below).

We perform the fit on the F814W luminosity function at $R_{\mathrm{eff}}<R<4R_{\mathrm{eff}}$, with 50 magnitude bins between $25.50$ and $27.85$ mag. The best fit is obtained for a TRGB magnitude of $m_{F814W, \rm TRGB}=27.47^{+0.04}_{-0.02} \pm 0.16$, where the first error is statistical and the second is systematic. The reduced $\chi^2$ of the best fit is $0.98$. The statistical uncertainty encompasses 68\,\% of the likelihood function (marginalized over $c$) from the forward modeling fitting. The main source of systematic uncertainty is that we strongly rely on the results of the artificial star tests due to our relatively low S/N ratio (compared to studies of galaxies that are much closer). The artificial stars show that both the magnitude bias and the photometric scatter are $0.15 \ \mathrm{mag}$ near the TRGB location. This bias and scatter are taken into account in our forward modeling, but we conservatively assign a $0.15 \ \mathrm{mag}$ systematic uncertainty to our TRGB measurement. Additionally, we find that varying the fitting range and the quality cuts to the photometry lead to $\pm 0.04$  mag variations in the measured TRGB magnitude. The final systematic uncertainty of $0.16 \ \mathrm{mag}$ is the quadratic sum of these two contributions.

The best-fitting model is shown in Figure \ref{fig:forward-modeling}, both before (``intrinsic model") and after applying the photometric errors.
The derived TRGB magnitude is fainter than that determined from the edge detection (see above). The reason is the photometric scatter, which has a larger effect on the apparent location of the TRGB than the photometric bias: due to the steepness of the luminosity function the number of stars that scatter from faint to bright magnitudes is much larger than that scattering from bright to faint magnitudes.

The absolute magnitude of the TRGB in the $F814W$-band is taken from the calibration of Rizzi et al.\ (\citeyear{2007ApJ...661..815R}). Their zero-point calibration of the TRGB is color-dependent, and accounts for shifts in the TRGB location due to variation in metallicity and age:
\begin{equation}
M_{F814W}^{\mathrm{ACS}} = -4.06+0.20[(F606W-F814W)-1.23]
\end{equation}
Using a TRGB median color index of $1.35 \ \mathrm{mag}$, we derive a TRGB absolute magnitude of $M_{F814W} = -4.036 \ \mathrm{mag}$. We adopt the systematic uncertainty on the zero-point calibration from Rizzi et al. (\citeyear{2007ApJ...661..815R}) of $0.07 \ \mathrm{mag}$ which is the standard in the field (see, e.g., \citealt{2017AJ....154...51M}). Adding this uncertainty in quadrature leads to a final distance modulus of $31.50^{+0.04+0.16+0.07}_{-(0.02+0.16+0.07)} = 31.50 \pm 0.18 \ \mathrm{mag}$ and a distance of $20.0 \pm 1.6 \ \mathrm{Mpc}$.

\section{Discussion} \label{sec:discussion}

In this \textit{Letter} we have used deep HST/ACS imaging to determine a tip of the red giant branch distance to one of the two dark matter-deficient galaxies in the field of NGC\,1052, NGC\,1052-DF4. The new data, acquired in Cycle 26 thanks to the mid-cycle process, comprises 7 orbits of $F814W$ and 3 orbits of $F606W$ imaging. Together with the previously-obtained single orbit imaging in both bands the data are just deep enough to allow detection of individual stars fainter than the TRGB (see Figure \ref{fig:photometry}).

Our best estimate for the distance is $20.0 \pm 1.6$\,Mpc. This result is consistent with the SBF distance ($19.9 \pm 2.8 \ \mathrm{Mpc}$) and the TRGB lower limit ($>9.7 \ \mathrm{Mpc}$) presented in \citet{2018ApJ...868...96C} for this galaxy.  It is also consistent with the calibration-free SBF distance of $18.7 \pm 1.7$\,Mpc  derived in \citet{2018ApJ...864L..18V} for NGC\,1052-DF2, with the globular cluster-calibrated SBF distance of $20.4\pm 2.0$\,Mpc to that galaxy derived by Blakeslee \& Cantiello (\citeyear{2018RNAAS...2c.146B}), and with the distance of the elliptical galaxy NGC\,1052 (19.2\,Mpc -- 20.6\,Mpc; \citealt{2001ApJ...546..681T}; \citealt{2013AJ....146...86T}). Our results do not agree with the distance derived by \citet{2019ApJ...880L..11M}, who found $m_{F814W,\rm TRGB}=26.7\pm 0.1$ and $D_{\rm TRGB}=14.2 \pm 0.7 \ \mathrm{Mpc}$ from the same data that were previously analyzed by \citet{2018ApJ...868...96C} and \citet{2018ApJ...864L..18V}. The best-fit when forcing the TRGB magnitude to the Monelli \& Trujillo (\citeyear{2019ApJ...880L..11M}) value is shown by the broken line in Figure \ref{fig:forward-modeling}. All previous studies used the 1+1 orbits in $F814W$ and $F606W$ from program GO-14644 (PI: van Dokkum). We suggest that a combination of AGB stars and photometrically-contaminated giants in these shallow data were incorrectly interpreted as the TRGB by \citet{2019ApJ...880L..11M}. Based on the near-identical CMDs of NGC\,1052-DF4 and NGC\,1052-DF2 we infer that this also happened in the \citet{2019MNRAS.486.1192T} analysis of NGC\,1052-DF2 \citep[for more discussion see][]{2018ApJ...864L..18V}.

Our analysis confirms the unusual nature of NGC\,1052-DF4. The spectroscopically-confirmed globular clusters of NGC\,1052-DF4 rival the most luminous globular clusters in the Milky Way, and its kinematics are consistent with the galaxy having little or no dark matter. The velocity dispersion of the galaxy is $\sigma_{\mathrm{intr}}=4.2^{+4.4}_{-2.2} \ \mathrm{kms}^{-1}$ and the expected value from the stars alone is $\sigma_{\mathrm{stars}} \approx 7 \ \mathrm{km s}^{-1}$ \citep[see][]{2019ApJ...874L...5V}. High resolution spectroscopy of the diffuse light of NGC\,1052-DF4 may provide a more accurate measurement of the kinematics in the future.

Our analysis of the new data places NGC\,1052-DF2 and NGC\,1052-DF4 at a distance that is consistent, within the errors, with the distance to NGC\,1052 itself, at $\approx 20$\,Mpc. \footnote{The distances of other group members are highly uncertain; see van Dokkum et al.\ (\citeyear{2019RNAAS...3...29V}) and Monelli \& Trujillo (\citeyear{2019ApJ...880L..11M}) for a discussion.}
The fact that the group is dominated by an elliptical galaxy is likely important in the context of understanding their formation. Proposed formation scenarios include assembly in the chaotic gas-rich environment of the assembling central galaxy (\citealt{2018Natur.555..629V}), severe tidal stripping (\citealt{2018MNRAS.480L.106O}; \citealt{2019A&A...624L...6M}), formation in a QSO outflow from NGC\,1052's black hole (\citealt{1998MNRAS.298..577N}) and high velocity collisions of gas-rich dwarfs at early epochs (\citealt{2019MNRAS.tmpL.105S}). At minimum, the presence of an overdensity likely influenced the probability of the formation of these galaxies. We note that the confirmation of the distance also confirms the presence of extremely luminous globular clusters in NGC\,1052-DF2 and NGC\,1052-DF4, and that any formation scenario should account for these objects.

Looking ahead, we also expect a distance measurement for NGC\,1052-DF2 in the near future. This galaxy will be imaged with {\em HST} in Cycle 27 for 38 orbits, bringing the total exposure time on that galaxy to 20 orbits in $F814W$ and 20 in $F606W$. These data will be taken in three epochs to help identify the AGB/RGB boundary using stellar variability.

\acknowledgments
We thank the anonymous referee for a very helpful and constructive review. 
Support from NASA STScI grants HST-GO-14644 and HST-GO-15695 is gratefully acknowledged. AJR was supported by National Science Foundation grant AST-1616710 and as a Research Corporation for Science Advancement Cottrell Scholar.


\end{document}